\documentclass[preprint,12pt]{elsarticle}

\usepackage{amssymb,wasysym}
\usepackage{amsfonts}
\usepackage{epstopdf}
\usepackage{wrapfig}

\newcommand{\beq}{\begin{equation}}
\newcommand{\eeq}{\end{equation}}
\newcommand{\ben}{\begin{eqnarray*}}
\newcommand{\een}{\end{eqnarray*}}

\def\eq#1{{(\ref{#1})}}

\journal{Progress in Particle and Nuclear Physics}

\begin{document}

\begin{frontmatter}



\title{The Chiral Magnetic Effect\\ and Anomaly-Induced Transport}


\author{Dmitri E. Kharzeev\footnote{e-mail:
{\tt dmitri.kharzeev@stonybrook.edu}}}

\address{{\it Department of Physics and Astronomy, Stony Brook University,} \\
{\it Stony Brook, New York 11794-3800 }\\[0.1in]
{\it Department of Physics, Brookhaven National Laboratory,} \\
{\it Upton, New York 11973-5000 }\\[0.1in]}

\begin{abstract}
The Chiral Magnetic Effect (CME) is the phenomenon of electric charge separation along the external magnetic field that is induced by the 
chirality imbalance. The CME is a macroscopic quantum effect - it is a manifestation of the chiral anomaly creating a collective motion in Dirac sea. 
Because the chirality imbalance is related to the global topology of gauge fields, the CME current is topologically protected and hence 
{\it non-dissipative} even in the presence of strong interactions. As a result, the CME and related quantum phenomena affect the hydrodynamical 
and transport behavior of systems possessing chiral fermions, from the quark-gluon plasma to chiral materials. The goal of the present review is to provide 
an elementary introduction 
into the main ideas underlying the physics of CME, a historical perspective, and a guide to the rapidly growing literature on this topic.
\end{abstract}




\end{frontmatter}

\newpage
\section{The Chiral Magnetic Effect in a nutshell}
\label{}
\vskip0.2cm
\hskip 3cm {\it ``The power of intuitive understanding will protect you} 

\hskip 3.2cm {\it from harm until the end of your days."} 
\vskip0.2cm
\hskip 10cm {\it Lao Tzu}

\vskip0.3cm
The goal of this review is to provide an easily accessible introduction into a new and rapidly developing field -- the macroscopic manifestations of quantum anomalies in collective dynamics of systems possessing chiral fermions.  In this section, we outline the main ingredients of the Chiral Magnetic Effect (CME), an exemplar macroscopic phenomenon stemming from the chiral anomaly; the subsequent sections provide details and references to the literature. For a more detailed exposure, the reader is referred to a recent volume \cite{Kharzeev:2012ph}.

\subsection{Magnetic field as a coherent probe of strongly interacting matter}
The magnetic field is a crucial ingredient of the CME as it breaks the rotational invariance and  
creates a preferred orientation for the spins of the fermions. The use of magnetic field in elucidating topological effects in condensed matter systems is widespread and has led to many breakthroughs, including the discovery of the Quantum Hall Effect (QHE)\footnote{As we will see later, the CME can be considered as an analog of QHE in $(3+1)$ dimensions.}. 
\begin{figure}[t]
	\centering
	\includegraphics[width=9cm]{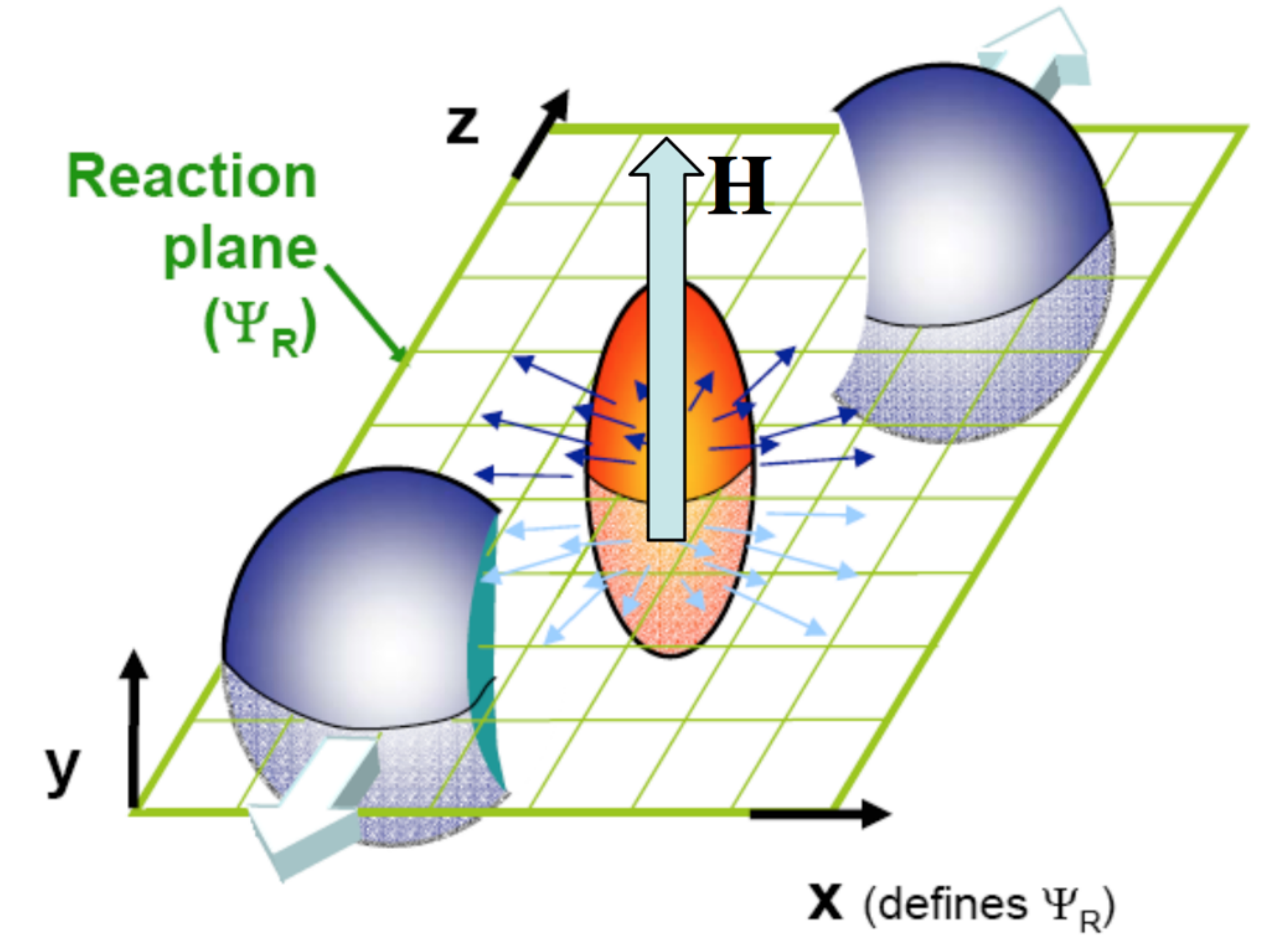}
		\caption{Collision of relativistic heavy ions produces a hot QCD matter penetrated by the flux of a strong magnetic field. \label{mag_field}}
\end{figure}
However in the physics of quark-gluon matter the use of magnetic fields is relatively new, and deserves a comment.  
\vskip0.3cm
Hard electromagnetic probes have proved crucial for understanding the strongly interacting matter -- the discovery of Bjorken scaling in deep-inelastic scattering has established quarks as constituents of the proton, and has opened the path towards the Quantum Chromo-Dynamics (QCD) and the asymptotic freedom. The subsequent development of QCD has led to the understanding of the role of extended field configurations in the non-perturbative dynamics of the theory, including spontaneous breaking of chiral symmetry and confinement. The hard probes are not well suited for the study of extended gluon field configurations because of the mismatch in scales -- a microscope is not the best tool if we are to distinguish between an elephant and a mouse. 
\vskip0.3cm
The challenge of collective dynamics in QCD thus calls for the study of response of strongly interacting matter to intense {\it coherent}  electromagnetic fields. 
Experimental access to the study of QCD plasma in very intense magnetic fields with magnitude $eB \sim 10\ m_{\pi}^2$ (or $\sim10^{18}\ \mathrm{G}$) \cite{Kharzeev:2007jp,Skokov:2009qp} is provided by the collisions of relativistic heavy ions. At nonzero impact parameter, these collisions create the magnetic field that is aligned, on the average, perpendicular to the reaction plane, see Fig. \ref{mag_field}.
Somewhat weaker magnetic fields $\sim 10^{15}\ \mathrm{G}$ exist in magnetars, where it may affect the properties of cold dense nuclear or quark matter. 

\subsection{The chiral anomaly and Dirac sea}\label{ansect}
\vskip0.2cm
\hskip 6cm {\it ``Become totally empty} 

\hskip 6cm {\it quiet the restlessness of the mind} 

\hskip 6cm {\it only then will you witness everything}

\hskip 6cm {\it unfolding from emptiness."}

\hskip 10cm {\it Lao Tzu}

\vskip0.3cm

\begin{figure}[t]
	\centering
	\includegraphics[width=10cm]{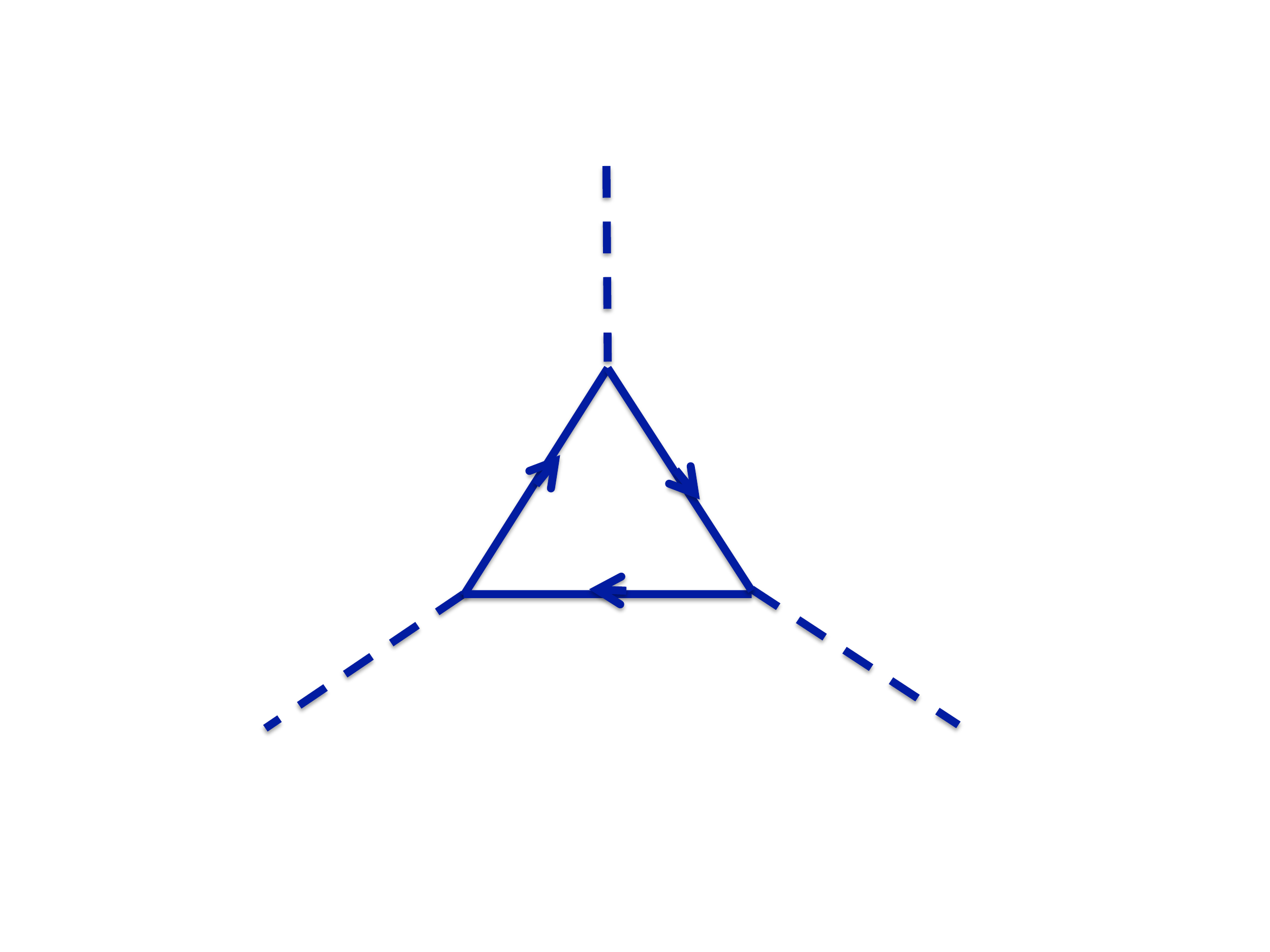}
		\caption{The triangle diagram involving the charged chiral fermion loop, a composite axial or dilatational current, and two external gauge fields.\label{triangle}}
\end{figure}

The term ``quantum anomaly" refers to the situation when a classical symmetry of the theory is broken by quantum effects. For example, 
QCD in the chiral limit of massless quarks possesses the chiral and scale invariances leading, by Noether's theorem, to the conservation of axial and dilatational currents. 
However, the regularization of quark triangle diagrams in background gauge fields (see Fig. \ref{triangle})  induces non-conservation of both currents, breaking explicitly 
the flavor-singlet $U_A(1)$ chiral symmetry and the scale symmetry. The resulting chiral\footnote{We will use the terms "chiral anomaly" and "axial anomaly" intermittently.} \cite{Adler:1969gk,Bell:1969ts} and scale \cite{Callan:1970yg,Symanzik:1970rt,Coleman:1970je,Ellis:1970yd,Collins:1976yq} anomalies imply spectacular consequences for the hadron spectrum --
the latter anomaly is responsible for the masses of all hadrons, and the former -- for the unexpectedly large mass of the pseudoscalar $\eta'$ meson which would be a Goldstone boson in the absence of the $U_A(1)$ anomaly.

The fermions with different chiralities contribute to the triangle diagram of Fig. \ref{triangle} with 
opposite signs -- as a result, the anomaly is absent for the vector current $J_V\equiv J_L+J_R$, and the electric charge is conserved. 
On the other hand, for the axial current $J_A\equiv -J_L+J_R$ interacting with an Abelian gauge field it leads to
\beq\label{triangle_an}
\partial_\mu J^\mu_A = {e^2\over 2\pi^2} \vec E\cdot \vec B\quad,
\eeq
where $e$ is the charge of the fermion (we implicilty assume a sum over charged fermion species), and $\vec E$ and $\vec B$ are the electric and magnetic fields. 
\vskip0.3cm

The key observation that will lead us to CME is that in the presence of chirality imbalance (which can be conveniently described by the chiral chemical potential\footnote{It is important to keep in mind that unlike the ordinary chemical potentials, the chiral chemical potential does not correspond to a conserved quantity -- the corresponding chiral charge is not conserved because of the chiral anomaly (even though a conserved global chiral charge can still be constructed with the help of Chern-Simons current density). We will see that this fact is crucial for the existence of CME.} $2 \mu_5 \equiv \mu_R - \mu_L$) the cancellation between 
the contributions of left- and right-handed fermions to the divergence of vector current is not complete. Note however that this  does not lead to 
the global non-conservation of electric charge $Q$, since the resulting total charge flow through a closed boundary $S$ surrounding the domain with $\mu_5 \neq 0$ vanishes: 
\beq
\frac{d Q}{d t} = - \oiint_S \vec j_V\ d \vec{S} = 0.
\eeq
\vskip0.3cm

While we now understand how the axial anomaly may generate electromagnetic current, it is still not clear how this purely quantum effect can operate on macroscopic scales. Indeed, a macroscopic phenomenon involves a very large number of quanta, whereas this is apparently not so 
for the effect described by Fig. \ref{triangle}. To resolve this apparent contradiction, we have to look deeper into the origin of quantum anomalies. As pointed out by Gribov \cite{Gribov:1981ku}, the source of anomalies can be traced back to the collective motion of particles with arbitrarily large momenta in the vacuum. This collective motion defies any UV cutoff that we may try to impose and
  "transfers the axial charge and the energy-momentum from the world with infinitely large momenta to our world of finite momenta" \cite{Gribov:1981ku}. 
  This is very different from the ``usual" quantum phenomena where we can always choose a UV cutoff that is sufficiently large to isolate the world of small momenta and large distances from the effects of large momentum, short distance quantum fluctuations.
  \vskip0.3cm
 Let us illustrate this statement for the case of axial anomaly by considering the Dirac sea of massless fermions. In the absence of external fields (or parity-odd interactions), the chirality is conserved and there are two disconnected Fermi surfaces of left- and right-handed fermions.  Let us now turn on adiabatically the external classical fields capable of changing the chirality of fermions -- e.g. the parallel electric $\vec{E}$ and magnetic $\vec{B}$ fields.  This field configuration will skew the balance between the Fermi surfaces of left- and right-handed fermions in the Dirac sea, transforming left-handed antiparticles into right-handed particles, or vice-versa, depending on the sign of the product $\vec{E}\cdot\vec{B}$, see (\ref{triangle_an}).
 \vskip0.3cm
 The mechanism of the collective flow of chirality can be described as follows \cite{Nielsen:1983rb,Witten:1984eb}:   
  the presence of magnetic field $B$ aligns the spins of the positive (negative) fermions in the direction parallel (anti-parallel) to $\vec{B}$. In the electric field $E$ the positive fermions will experience the force $e E$ and will move along $\vec{E}$; therefore their spins will have a positive projection on momentum, and we are dealing with the right fermions. Likewise, the negative fermions will be left-handed. 
After time $t$, the positive (right) fermions will increase their Fermi momentum to $p^F_R = e E t$, and the negative (left) will have their Fermi momentum decreased to $p^F_L = - p^F_R$. The one-dimensional density of states along the axis $z$ (that we choose parallel to the direction of fields $\vec E$ and $\vec B$) is given by $dN_R / dz = p^F_R / 2 \pi$. In the transverse direction, the motion of fermions is quantized as they populate Landau levels in the magnetic field. The transverse density of Landau levels is $d^2 N_R/ dx dy = e B/ 2 \pi$. Therefore the density of right fermions increases per unit time as 
\beq
\frac{d^4 N_R}{dt\ dV} = \frac{e^2}{(2 \pi)^2}\ \vec{E} \cdot \vec{B}.
\eeq
The density of left fermions decreases with the same rate, $d^4 N_L / dt\ dV = -  d^4 N_R / dt\ dV$. The local rate of chirality $Q_5 = N_R - N_L$ generation is thus
\beq\label{chirchange}
\frac{d^4 Q_5}{dt\ dV} = \frac{e^2}{2 \pi^2}\ \vec{E} \cdot \vec{B},
\eeq 
in accord with (\ref{triangle_an}).
\vskip0.3cm
The quantity on the r.h.s. is the density of topological Chern-Pontryagin charge; its integral over four-dimensional space 
\beq\label{topcharge}
q[A] = {\frac{e^2}{8 \pi^2}} \int d^4 x\ F^{\mu\nu} \tilde{F}_{\mu\nu};
\eeq
reveals the  topological class to which the vector potential $A$ belongs. It has to be integer, just as the difference between the numbers of right- and left-handed fermions. 
The relation (\ref{chirchange}) thus expresses the deep connection between the axial anomaly and the topology of classical gauge fields that is formally expressed by the Atiyah-Singer theorem for the index of Dirac operator \cite{Atiyah:1963zz,Atiyah:1984tf}.
\vskip0.3cm
Having a  classical field with an infinite number of quanta is important here since the picture described above involves changing the momenta of an infinite number of particles, and "a finite number of photons is not able to change the momenta of an infinite number of particles" \cite{Gribov:1981ku}. This feature of the anomaly provides an intuitive explanation of the absence of perturbative quantum corrections to the axial anomaly that can be established formally through the renormalization group arguments \cite{Zee:1972zt,Lowenstein:1973fa,Adler:2004qt}. As will be discussed below, the (electromagnetic) axial anomaly is robust even when the coupling constant that determines the strength of (non-electromagnetic) interactions among the fermions becomes infinitely large. This is because the anomaly relation is topologically protected -- the net chirality is related to the topological class of the gauge field that is determined at large distances, so the local interactions cannot change it. 

\vskip0.3cm
The flow of chirality, as the derivation above reveals, is accompanied by the collective motion of particles at all momenta, including the momenta around the UV cutoff scale $\Lambda_{UV}$ that we may attempt to introduce. Therefore our world of particles with finite momenta $p < \Lambda_{UV}$ cannot be isolated from particles with arbitrarily high momenta, and this is the essence of quantum anomaly.

\subsection{The Chiral Magnetic Effect}\label{cme_sect}

\vskip0.2cm
We now have all the ingredients needed to discuss the Chiral Magnetic Effect (CME) -- the phenomenon of electric charge separation along the external magnetic field induced by the chirality imbalance. To begin, we will follow the derivation from \cite{Kharzeev:2009fn} as it is simple and highlights the connection of CME to topology of gauge fields\footnote{For a chronologically ordered presentation, see next section.}. 
Let us couple QCD to electromagnetism; the resulting theory possesses $SU(3) \times U(1)$ gauge symmetry:
$$
{\cal L}_{\rm QCD+QED} =  -{1 \over 4} G^{\mu\nu}_{\alpha}G_{\alpha \mu\nu}  + \sum_f \bar{\psi}_f \left[ i \gamma^{\mu} 
(\partial_{\mu} - i g A_{\alpha \mu} t_{\alpha} -  i q_f A_{\mu}) -  m_f \right] 
\psi_f  - 
$$
\beq\label{qcd+qed}
- {\theta \over 32 \pi^2}  g^2 G^{\mu\nu}_{\alpha} \tilde{G}_{\alpha \mu\nu} - \frac{1}{4}F^{\mu\nu}F_{\mu\nu},
\eeq 
where $A_{\mu}$ and $F_{\mu\nu}$ are the electromagnetic vector potential and the field strength tensor, $A_{\alpha \mu}$ and $G_{\alpha \mu\nu}$ are the corresponding quantities for the gluon fields, $q_f$ and $m_f$ are the electric charges and masses of the quarks, $g$ is the strong coupling, and  we have allowed for the P- and CP-odd $\theta$-term.  
\vskip0.3cm
Let us discuss the electromagnetic sector of the theory  \eq{qcd+qed}. Electromagnetic fields will couple to the electromagnetic currents $J_\mu = \sum_f  q_f \bar{\psi}_f \gamma_\mu \psi_f$.  
In addition, the quark loop of Fig.\ref{triangle} will induce the coupling of $F \tilde{F}$ to the QCD topological charge density which is the non-Abelian extension of (\ref{topcharge}) and is induced by instantons \cite{Belavin:1975fg}. We will introduce an effective pseudo-scalar field $\theta = \theta(\vec x, t)$ and write down the resulting effective Lagrangian as
\beq\label{MCS}
{\cal L}_{\rm MCS} = - \frac{1}{4}F^{\mu\nu}F_{\mu\nu} - A_\mu J^\mu - \frac{c}{4}\ \theta \tilde{F^{\mu\nu}}F_{\mu\nu},
\eeq
where 
\beq\label{coef}
c = \sum_f q_f^2 e^2 / (2\pi^2). 
\eeq

This is the Lagrangian of Maxwell-Chern-Simons, or axion, electrodynamics that has been introduced previously in \cite{Wilczek:1987mv,Carroll:1989vb,Sikivie:1984yz}. 
If $\theta$ is a constant, then the last term in \eq{MCS} is a full divergence 
\beq\label{an_ab}
\tilde{F^{\mu\nu}} F_{\mu\nu} = \partial_\mu K^\mu
\eeq
of the Abelian Chern-Simons current 
\beq\label{topdiv1}
K^{\mu} = \epsilon^{\mu\nu\rho\sigma} A_{\nu} F_{\rho\sigma}.
\eeq
Being a full divergence, this term 
does not affect the equations of motion and thus does not affect the electrodynamics of charges.

\vskip0.3cm

The situation is different if the field $\theta = \theta(\vec x, t)$ varies in space-time.      
Indeed, in this case we have
\beq\label{fullder}
\theta \tilde{F^{\mu\nu}} F_{\mu\nu} = \theta \partial_\mu K^\mu = \partial_\mu\left[\theta K^{\mu}\right] - \partial_\mu \theta  K^{\mu}.
\eeq
The first term on r.h.s. is again a full derivative and can be omitted; introducing notation
\beq
P_\mu = \partial_\mu \theta = ( M, \vec P )
\eeq
we can re-write the Lagrangian \eq{MCS} in the following form:
\beq\label{CS}
 {\cal L}_{\rm MCS} = - \frac{1}{4}F^{\mu\nu}F_{\mu\nu} - A_\mu J^\mu + \frac{c}{4} \ P_\mu K^\mu
\eeq
Since $\theta$ is a pseudo-scalar field, $P_\mu$ is a pseudo-vector; as is clear from   \eq{CS}, 
it plays a role of the potential coupling to the Chern-Simons current \eq{topdiv1}. However, unlike the vector potential $A_\mu$, $P_\mu$ is not a dynamical variable and is a pseudo-vector that is fixed by the dynamics of chiral charge -- in our case, determined by the fluctuations of topological charge in QCD.
\vskip0.3cm
In $(3+1)$ space-time dimensions, the pseudo-vector $P_\mu$ selects a direction in space-time and thus breaks the Lorentz and rotational invariance \cite{Carroll:1989vb}: the temporal component $M$ breaks the invariance w.r.t. Lorentz boosts, while the spatial component $\vec P$ picks a certain direction in space. On the other hand, in $(2 + 1)$ dimensions there is no need for the spatial component $\vec P$ since the Chern-Simons current \eq{topdiv1} in this case reduces to the pseudo-scalar quantity $\epsilon^{\nu\rho\sigma} A_{\nu} F_{\rho\sigma}$, so the last term in \eq{CS} takes the form
\beq\label{2+1}
\Delta {\cal L} = c\ M \epsilon^{\nu\rho\sigma} A_{\nu} F_{\rho\sigma}.
\eeq
This term is Lorentz-invariant although it still breaks parity. 
In other words, in $(2+1)$ dimensions the vector $\vec P$ can be chosen as a 3-vector pointing in the direction of an "extra dimension" orthogonal to the plane of the two spatial dimensions. When added to the Maxwell action, \eq{2+1} generates a mass of the photon which thus becomes "topologically massive" \cite{Deser:1981wh}. This illustrates an important difference between the roles played by Chern-Simons term in even and odd number of space-time dimensions.   
\vskip0.3cm
The equation of motion derived for the sum of the Lagrangian \eq{2+1} and the term $A_\mu J^\mu$ is $J^\mu \sim \epsilon^{\mu\rho\sigma} F_{\rho\sigma}$. Integrating the temporal component of this equation over the spatial plane, we find that in $(2+1)$ dimensions the Chern-Simons term endows charged particles with magnetic flux -- this is a celebrated phenomenon \cite{frankbook} that is an essential ingredient of the Quantum Hall Effect. 
 
 \vskip0.3cm
  
   
Let us now write down the Euler-Lagrange equations of motion that follow from the Lagrangian \eq{CS}  
(Maxwell-Chern-Simons equations) in $(3+1)$ dimensions:
\beq
\partial_\mu F^{\mu\nu} = J^\nu - P_\mu \tilde{F}^{\mu\nu};
\eeq
the first pair of Maxwell equations (which is a consequence of the fact that the fields are expressed through the vector potential) is not modified:
\beq
\partial_\mu \tilde{F}^{\mu\nu} = 0.
\eeq   
It is convenient to write down these equations also in terms of the electric $\vec E$ and magnetic $\vec B$ fields:
\beq\label{MCS1}
\vec{\nabla}\times \vec{B} - \frac{\partial \vec{E}}{\partial t} = \vec J + c \left(M \vec{B} - \vec{P} \times \vec{E}\right), 
\eeq
\beq\label{MCS2}
\vec{\nabla}\cdot \vec{E} = \rho + c \vec{P} \cdot \vec{B},
\eeq
\beq\label{MCS3}
\vec{\nabla}\times \vec{E} +  \frac{\partial \vec{B}}{\partial t} = 0,
\eeq
\beq\label{MCS4}
\vec{\nabla}\cdot \vec{B} = 0,
\eeq
where $(\rho, \vec J)$ are the electric charge and current densities.
One can see that the presence of Chern-Simons term leads to essential modifications of the Maxwell theory. 
\vskip0.3cm
Let us consider the case when $| \vec{P} | =0$ but  $\dot{\theta} \neq 0$. We will introduce an external magnetic field $\vec B$ with $\vec{\nabla}\times \vec{B} = 0$, and assume that no external electric field is present.  In this case we immediately get from \eq{MCS1} that there is an induced current 
 \beq\label{chimag}
 \vec{J} = - c\ M\ \vec{B} = - \frac{e^2}{2 \pi^2}\ \dot\theta \vec{B},
 \eeq
 where $\dot\theta$ has to be identified with the chiral chemical potential $\mu_5$, see \cite{Fukushima:2008xe} and \cite{Kharzeev:2009fn}. Similarly to the absence of perturbative corrections to the axial anomaly discussed above, there are no corrections to (\ref{chimag}) \cite{Hou:2011ze}, on the operator level.
 This does not mean that the expectation value of the current cannot be renormalized, see \cite{Gorbar:2009bm,Fukushima:2010zza,Gorbar:2010kc,Gorbar:2013upa,Gorbar:2013uga} for the studies of this topic. 
 \vskip0.3cm
 
 The same result can be obtained \cite{Fukushima:2008xe} by explicitly evaluating the thermodynamic potential $\Omega$ for charged chiral fermions in magnetic field at finite chiral chemical potential $\mu_5$, 
\beq
 \Omega = \frac{\vert eB \vert}{2\pi} \sum_{s=\pm} \sum_{n=0}^{\infty}
\alpha_{n,s}
\int_{-\infty}^{\infty} \frac{\mathrm{d} p_3}{2\pi}
\Bigl [ \omega_{p, s} 
+ T \sum_\pm \log (1 + e^{-\beta (\omega_{p, s} \pm \mu) })
\Bigr ],
\label{eq:thermopotmag}
\eeq
where $n$ is a sum over Landau levels, $s$ is a sum over spin projections and the
dispersion relation is given by
\begin{equation}
 \omega_{p,s}^2 = \left[
\mathrm{sgn}(p_3)  
(p_3^2 + 2 \vert  e B \vert n)^{1/2} +
s \mu_5  \right]^2 + m^2.
\label{eq:thermopotmagdispers}
\end{equation}
The differentiation of this potential w.r.t. the vector potential $A_3$ yields the density of current flowing along the direction of magnetic field
\begin{equation}
 j_3 = \left. \frac{\partial \Omega}{\partial A_3}
\right \vert_{A_3=0}
\end{equation}
and can be done by noting that $\partial/\partial A_3 = e d/d p_3$ from gauge invariance.  Performing this operation on \eq{eq:thermopotmag}
produces a boundary term and clearly demonstrates the relation of \eq{chimag} to the spectral flow in the Dirac sea \cite{Fukushima:2008xe}, which as we discussed above is the nature of axial anomaly. A very transparent picture of CME illuminating the role of anomaly emerges when one uses a dimensional reduction appropriate for strong magnetic field, as explained in \cite{Basar:2010zd}.
 \vskip0.3cm
 
  \subsection{CME current as a non-dissipative phenomenon}
 
 The absence of CME in conventional Maxwell electrodynamics follows already from the symmetry considerations -- indeed, the magnetic field is a (parity-even) pseudo-vector, and the electric current is a (parity-odd) vector. Therefore, CME signals the violation of parity -- indeed, as we discussed above, its presence requires the asymmetry between the left and right fermions.
 \vskip0.3cm
 Another unusual and very important property of the relation \eq{chimag} is that the ``chiral magnetic conductivity"  $(e^2/2 \pi^2)\ \dot\theta$ is even under time reversal $\cal T$. Indeed, both the electric current on the l.h.s. of \eq{chimag} and the magnetic field on the r.h.s. are ${\cal T}$-odd quantities. One can also see this directly since $\theta$ is a ${\cal T}$-odd ``axion" field, and differentiation w.r.t. time yields a ${\cal T}$-even quantity. 
 This is a highly unusual property for a conductivity. For example, the ``usual" electric conductivity $\sigma$ is ${\cal T}$-odd, as can be easily inferred from the Ohm's law $J^i = \sigma E^i$: the electric field is ${\cal T}$-even, whereas the electric current  $J^i$ is ${\cal T}$-odd. 
 This can also be illustrated by the Drude's formula for the ohmic conductivity that relates it to the mean free time between the collisions.
 On the other hand,  the (anomalous) quantum Hall conductance  is a ${\cal T}$-even quantity, as it is associated with a ${\cal T}$-odd magnetic field. 
 \vskip0.3cm
 The physical meaning of ${\cal T}$ invariance of transport coefficients is quite simple: ${\cal T}$-odd conductivities describe dissipative currents, whereas ${\cal T}$-even conductivities describe {\it non-dissipative} currents \cite{Kharzeev:2011ds}. 
 As an example, one can mention the London formula in superconductivity\footnote{I am grateful to V.I.Zakharov and L. Stodolsky who pointed this out to me.}:
 \beq\label{london}
 \vec J = - \frac{e^2 n}{m}\ \vec A,
 \eeq
 where $m$ is electron mass and $n$ is a phenomenological parameter describing the density of current carriers. Since both $\vec J$ and $\vec A$ are (${\cal T}$-odd) vectors, the ``conductivity" $e^2 n/m$ is ${\cal T}$-even, in line with the absence of dissipation for a superconducting current. The relation \eq{london} is of course not gauge-invariant, which is acceptable only because of the $U(1)$ breaking in the ground state of a superconductor due to the condensation of Cooper pairs, which is similar to the Higgs phenomenon. In contrast, the CME  \eq{chimag} relates two gauge invariant quantities $\vec J$ and $\vec B$ and requires only the imbalance of chiralities. 
 \vskip0.3cm
 The anomaly-induced currents are protected by topology and are thus of non-disipative nature; as such, they do not contribute to the entropy production. This principle, combined with the entropy current method \cite{Son:2009tf},  can be used to constrain the anomalous relatiivistic hydrodynamics and to evaluate analytically most of the anomalous transport coefficients that enter at second order in the gradient expansion \cite{Kharzeev:2011ds}.
 
 \vskip0.3cm
 
 
 Let us now discuss the nature of CME current from a different point of view. Consider the work done by the electric current \eq{chimag}; to obtain the work per unit time -- the power $P$ -- we multiply both sides of \eq{chimag} by 
 the (static) electric field $\vec{E}$ and integrate them over the volume (as before, we assume that $\theta$ does not depend on spatial coordinates):
 \beq\label{work}
 P = \int d^3 x\ \vec{J}\cdot \vec{E} = - \dot\theta\  \frac{e^2}{2 \pi^2}\  \int d^3 x\ \vec{E}\cdot \vec{B} = - \dot\theta\   \dot{Q}_5,
 \eeq
 where 
 \beq\label{topabel}
 {Q}_5 = \frac{e^2}{2 \pi^2}\  \int dt\ d^3 x\ \vec{E}\cdot \vec{B}
 \eeq
  is the chiral charge. 
 The meaning of the quantity on the r.h.s. of \eq{work} can be revealed with the help of the following well-known quantum-mechanical analogy. The $\theta$-vacuum wave function
\beq
|\theta\rangle = \sum_{Q_5} \exp(i\ \theta\ Q_5)\ |Q_5 \rangle
\eeq 
is analogous to the Bloch wave function of electron in a crystal, with $\theta$ playing the role of electron's quasi-momentum, 
and $Q_5$ -- the role of coordinates of atoms in the crystal. The derivative of the "momentum" $\dot\theta$ thus plays the role of the force and $Q_5$ -- of the dimensionless distance; $\dot{Q_5}$ is thus the velocity. The formula \eq{work} is therefore simply the classical expression
$$
{\rm Power} = {\rm Force} \times {\rm Velocity},
$$
with the force acting along the "extra dimension" of the chiral charge $Q_5$ \cite{Kharzeev:2009fn}. The fact that this power \eq{work} can be both positive or negative depending on the relative sign of $\vec E$ and $\vec B$ signals the absence of dissipation and thus of the arrow of time. 

\vskip0.cm

\section{CME: a historical perspective}\label{history}

\hskip 7cm {\it "All history becomes subjective;} 

\hskip 7cm {\it in other words there is properly} 

\hskip 7cm {\it no history, only biography." }\\

\hskip 7cm {\it Ralph Waldo Emerson}

\vskip0.3cm
During my early work on CME, I was not aware of many of the preceding important advances made by other people.
The purpose of this section is twofold: first, I would like to fill this gap and list the important preceding work as known to me at present; and second, to describe the path that has brought me and my collaborators to CME. I hope that this discussion may help to correct some misconceptions of the past and allow the readers to avoid them in the future. 
\vskip0.3cm

Let me first describe my path towards the CME. In 1995, when I was a postdoc at CERN Theory Division, I got interested in the problem of detection of the P and CP odd fluctuations in the QCD vacuum. While there was little doubt about the existence of such fluctuations (instantons \cite{Belavin:1975fg} provide a famous example), and these fluctuations were tied theoretically to many of the salient features of QCD, a direct experimental signature of their existence was lacking. During a discussion with Anatoly Efremov, who was visiting CERN from Dubna at the time, I learned about the idea of jet ``handedness" \cite{Nachtmann:1977ek,Efremov:1978qy,Efremov:1992pe} that allowed (at least, in principle) to reconstruct in experiment the chirality of the quark jet.  A very clean controlled source of quark-antiquark jets was provided by the decays of the $Z^0$ boson formed in $e^+e^-$ annihilation at the Large Electron-Positron collider that was operating at CERN at that time. When a quark enters a P- and CP-odd fluctuation of gluon field, its chirality changes -- and the correlation of chiralities of the quark and antiquark produced in the $Z^0$ decay can be used to isolate this effect. We estimated the magnitude of P- and CP-odd correlations in the fragmentation of quark and antiquark jets, and found a $\sim 1\%$ effect \cite{Efremov:1995ff}\footnote{For a recent extension of this idea, see \cite{Kang:2010qx}.}. Unfortunately, it appeared that the jet handedness correlation was not easy to measure, and no clear conclusion was reached on the existence of these correlations at LEP in spite of the hints reported by DELPHI Collaboration.

\vskip0.3cm
In 1997, I moved to the US to join the newly created RIKEN-BNL Center at Brookhaven that was founded and directed by T.D. Lee. I had several inspiring conversations with him that reaffirmed my interest in the fundamental symmetries of QCD. Moreover, during the inaugural RIKEN-BNL conference in 1997, Frank Wilczek gave a talk in which he discussed the possibility of spontaneous parity breaking in the color superconductor phase of QCD at large baryon density and low temperature. 
I decided to check whether a metastable P-odd states were possible at high temperature, in the regime that is accessible to heavy ion collisions. The Relativistic Heavy Ion Collider (RHIC) was under construction at Brookhaven, and there was already a lot of excitement about the program (that is in full swing at present). 

\vskip0.3cm
 Using a non-linear $\sigma$ model with axial anomaly to describe hadronic matter below the deconfinement phase transition, I found that close to the critical temperature, when the topological susceptibility decreased, the effective potential developed metastable minima corresponding to the P- and CP-odd domains where the $\theta$ angle of QCD was locally different from zero. I then gave a seminar on that work at BNL prior to submitting the paper for publication -- this was very fortunate, because after the seminar Rob Pisarski and Michel Tytgat found a big deficiency  in my arguments (some of the found ``minima" were instead unstable saddle points). Together, we have re-done the computation, identified the metastable phases, and pointed out the possibility to detect the P- and CP-odd domains through global observables constructed from charged pion momenta \cite{Kharzeev:1998kz}\footnote{Later we became aware of the paper by Morley and Schmidt \cite{Morley:1983wr} in which the authors hypothesized that the quark-gluon plasma could possess, globally, a non-zero $\theta$.}. 
 \vskip0.3cm
 
Soon after that paper appeared, the possibility to detect P-odd fluctuations in the quark-gluon plasma attracted attention of experimentalists -- Jack Sandweiss and his group at Yale, Sergei Voloshin at Wayne State, and Ron Longacre and Jim Thomas at Brookhaven decided to 
 perform a search for these phenomena within the STAR Collaboration at RHIC. Together with Rob Pisarski, in 1999 we proposed a number of global observables for heavy ion collisions \cite{Kharzeev:1999cz}, but simulations done by our experimental colleagues showed that using them for isolating the parity-odd domains was difficult \cite{Voloshin:2000xf}.
\vskip0.3cm
The commissioning of RHIC took place in 2000, and I spent most of the following three years working with Marzia Nardi and Genya Levin on a semi-classical  approach to multiparticle production \cite{Kharzeev:2000ph,Kharzeev:2001gp,Kharzeev:2001yq}  in high energy collisions (at present known as the ``KLN model") to describe the rapidly accumulating data. It was reassuring that within the general semi-classical QCD framework \cite{Gribov:1984tu,McLerran:1993ni}, the real-time lattice studies that we performed with Alex Krasnitz and Raju Venugopalan indicated the presence of strong chirality fluctuations at the early stages of heavy ion collisions  
\cite{Kharzeev:2001ev}. 
\vskip0.3cm

Among the early findings at RHIC was the observation of ``elliptic flow" -- the azimuthal anisotropy of produced hadrons \cite{Arsene:2004fa,Adcox:2004mh,Back:2004mh,Adams:2005dq}. The azimuthal distribution of charged hadrons produced in heavy ion collisions can be expanded in Fourier harmonics  in the following way: 
\beq\label{flow}
\frac{d N_{\pm}}{d\phi} \sim 1 + 2 v_1 \cos(\Delta \phi) + 2 v_2 \cos(2 \Delta \phi) +  ..., 
\eeq
where $\Delta \phi = \phi - \Psi_{RP}$ is the angle with respect to the reaction plane -- the plane which contains the impact parameter and beam momenta, see Fig. \ref{mag_field}. Note that a typical relativistic heavy ion collision produces several thousand hadrons, so the reaction plane can be reliably identified in each event.  The coefficients $v_1$ and $v_2$ measure the strength of so-called directed and elliptic flow. For symmetry reasons, in the collisions of identical nuclei, the directed flow $v_1$  vanishes at mid-rapidity (but not $\langle v_1^2 \rangle$ that measures the fluctuations).
\vskip0.3cm

 The elliptic flow $v_2$ signals the presence of the symmetry axis in the system of colliding ions that points perpendicular to the reaction plane. 
I realized that the existence of this symmetry axis defined by the angular momentum of the colliding ions was the crucial ingredient that could allow to detect 
the violation of parity -- similarly to the way the angular momentum of the $^{60}{\rm Co}$ was utilized to observe the asymmetry of $\beta$ decay in a classic experiment of C.-S. Wu following the ground-breaking idea of T.D. Lee and C.N. Yang. 
\vskip0.3cm

The key is the observation that a local P and CP-odd domain can be described as a region with a space-time dependent $\theta(\vec x, t)$ angle. Because of this, the domain can transfer energy and momentum  to quark--antiquark pairs, and ``can generate chirality not by flipping the spins of the quarks, but by inducing up--down asymmetry
(as measured with respect to the symmetry axis) in the production of quarks and antiquarks" \cite{Kharzeev:2004ey}. This asymmetry in the production of 
quarks and antiquarks generates the electric current along the direction of the angular momentum (and/or magnetic field), and results in the charge asymmetry with respect to the reaction plane; experimentally, one may detect it through the P-odd harmonics $a_+ = -a_- \equiv a$ in Eq. (\ref{asym}) that have the opposite signs and the magnitude of $\sim 1\%$ \cite{Kharzeev:2004ey}:
\beq\label{asym}
\frac{d N_{\pm}}{d\phi} \sim 1 \pm 2 a \sin(\Delta \phi) + ...  .
\eeq
\vskip0.3cm
Of course, since QCD does not violate P and CP globally (as we know e.g. from the measurements of the electric dipole moment of the neutron \cite{Baker:2006ts}), the average value $\langle \theta(\vec x, t) \rangle =0$. This means that the charge asymmetry should fluctuate event-by-event, and the signature of the effect is the dynamical fluctuations that exceed the statistical ones.  
\vskip0.3cm

\begin{figure}[t]
	\centering
	\includegraphics[width=9cm]{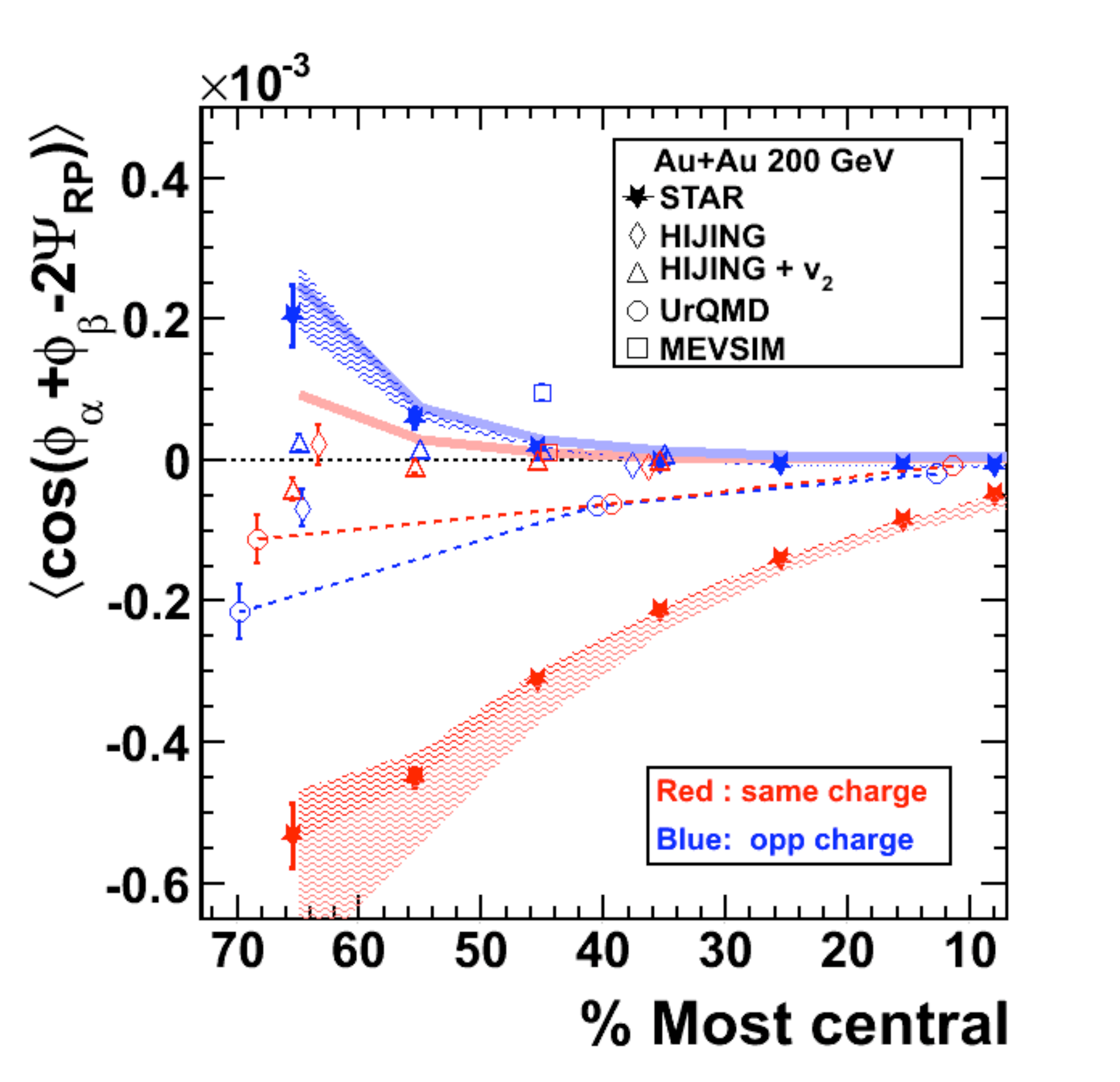}
	\caption{
The STAR Collaboration result on the charge-dependent azimuthal correlations in Au-Au collisions at $\sqrt{s} = 200$ GeV per nucleon pair at RHIC; from \cite{Abelev:2009uh}. Horizontal axis is the centrality of the collision (the fraction of the inclusive inelastic cross section); 
the average impact parameter decreases towards the right, as does the magnetic field.  Also shown are the predictions of various Monte-Carlo models of heavy ion collisions. 
 }
\label{simulations}
\end{figure}

In 2004, I presented my paper \cite{Kharzeev:2004ey} it to the experimental colleagues at RHIC, including Sergei Voloshin. Sergei called my story a ``fairy tale", and looked skeptical - but just in a couple of weeks, he came up with an ingenious way to look for the effect \cite{Voloshin:2004vk}. 
Let me briefly present the idea of \cite{Voloshin:2004vk}. The number of charged hadron tracks in a single event (although sufficient to determine the reaction plane) is not large enough to allow a statistically sound extraction of the coefficients $a_\pm$, so one has to sum over many events. However since there is no global violation of P and CP invariances in QCD, the sign of the charge asymmetry should fluctuate event by event and so when averaged over many events, $\left< a_+ \right> = \left< a_+ \right> =0$. The way out of this dilemma proposed  by Sergei was  to extract the cumulant 
$\left< a_\alpha a_\beta \right>$ (the indices $\alpha$ and $\beta$ denote the charge of hadrons) by measuring the expectation value of   $\left< \sin(\Delta \phi_\alpha) \sin(\Delta \phi_\beta) \right>$. 
The proposed in \cite{Voloshin:2004vk} variable 
\beq\label{variable}
\left< \cos(\phi_\alpha + \phi_\beta - 2 \Psi_{RP}) \right> = \left< \cos \Delta \phi_\alpha \cos \Delta \phi_\beta \right> - \left< \sin \Delta \phi_\alpha \sin \Delta \phi_\beta \right>
\eeq
has an added benefit of not being sensitive to the reaction plane--independent backgrounds that cancel out in \eq{variable}. The quantity \eq{variable} can be measured with a very high precision, and is directly sensitive to the parity--odd fluctuations. The price to pay however is that the observable itself is parity--even, and so one has to carefully examine all possible backgrounds.
\vskip0.3cm
In 2005, Ilya Selyuzhenkov on behalf of STAR Collaboration presented the preliminary result \cite{Selyuzhenkov:2005xa} on the measurement of \eq{variable}: there was a clear difference between the same-charge and opposite-charge cumulants that had a different sign, as expected! The period of excitement however ended a few months later when an unexpected effect was detected:
the cumulants appeared very sensitive to the type of hadrons (positive, negative, or combined charged) that was used to determine the reaction plane. Unless a physical explanation of this dependence existed, this signaled that the observed effect was not real and probably originated from some poorly understood systematics. I was asked whether this dependence could be explained, and spent three months trying to understand it -- but could not find any rational explanation of the phenomenon. The initial excitement had all but subsided, and the topic was rapidly fading into oblivion.
\vskip0.3cm
Fortunately, about a year later the re-analysis made by STAR in collaboration with Yannis Semertzidis and Vassily Dzhordzhadze at Brookhaven had revealed a coding bug responsible for the puzzling dependence -- once it was corrected, the result was qualitatively consistent with the theoretical expectations  \cite{Voloshin:2008jx}. The conclusive measurement of \eq{variable} was presented by STAR Collaboration in \cite{Abelev:2009uh,Abelev:2009ad}; one of the results is shown in Fig.\ref{simulations}. One can see that the same-charge and opposite-charge cumulants \eq{variable} differ in a way that is very significant statistically.  The predictions of various Monte Carlo models of heavy ion collisions are also shown in Fig.\ref{simulations}; these models (while successful in reproducing the global features of heavy ion collisions) fail in explaining the observed effect. Yet, it is too early to claim a victory since mundane backgrounds exist and are not excluded. The recent data have added a crucial information, and I believe that a definite conclusion can be reached in the near future -- but this is already the modern history described in the next section of this review.

\vskip0.3cm
On the theory front, it was clear that the arguments of \cite{Kharzeev:2004ey} on the separation of electric charge had to be strengthened.  
In the Fall of 2006, I attended a ``Quark confinement" conference on A\c{c}ores islands together with Eric Zhitnitsky. I told him about the unfolding story 
on P violation, and Eric got excited -- probably because he sensed the connection of the effect to the flow of axial current along the vortices in cold dense matter that he investigated with Dam Son \cite{Son:2004tq} and Max Metlitski \cite{Metlitski:2005pr}.
We began to collaborate, and in 2007 published a paper \cite{Kharzeev:2007tn} that clarified the role of magnetic field and vorticity in the ``charge separation effect". We used the space-time dependent $\theta$-angle to express the electric charge 
induced by the anomaly in the presence of vorticity $\Omega$ as 
\beq
J^0 =  N_c\ \frac{e \mu}{2 \pi^2}\cdot      \left(\vec{\nabla}\theta\cdot \vec{\Omega}\right),
\eeq
where $\mu$ is the chemical potential and $N_c$ is the number of colors; the sum over flavors is implicit here. 
Later, this phenomenon of electric charge separation induced by vorticity was described by the term ``Chiral Vortical Effect" (CVE) introduced in \cite{Kharzeev:2009fn}. The CVE was found within the fluid-gravity correspondence in 2008 \cite{Erdmenger:2008rm,Banerjee:2008th}, and in 2009 in the framework of 
relativistic hydrodynamics \cite{Son:2009tf}; its correlation with CME was discussed in \cite{Kharzeev:2010gr}.
\vskip0.3cm

\begin{figure}[t]
	\centering
	\includegraphics[width=10cm]{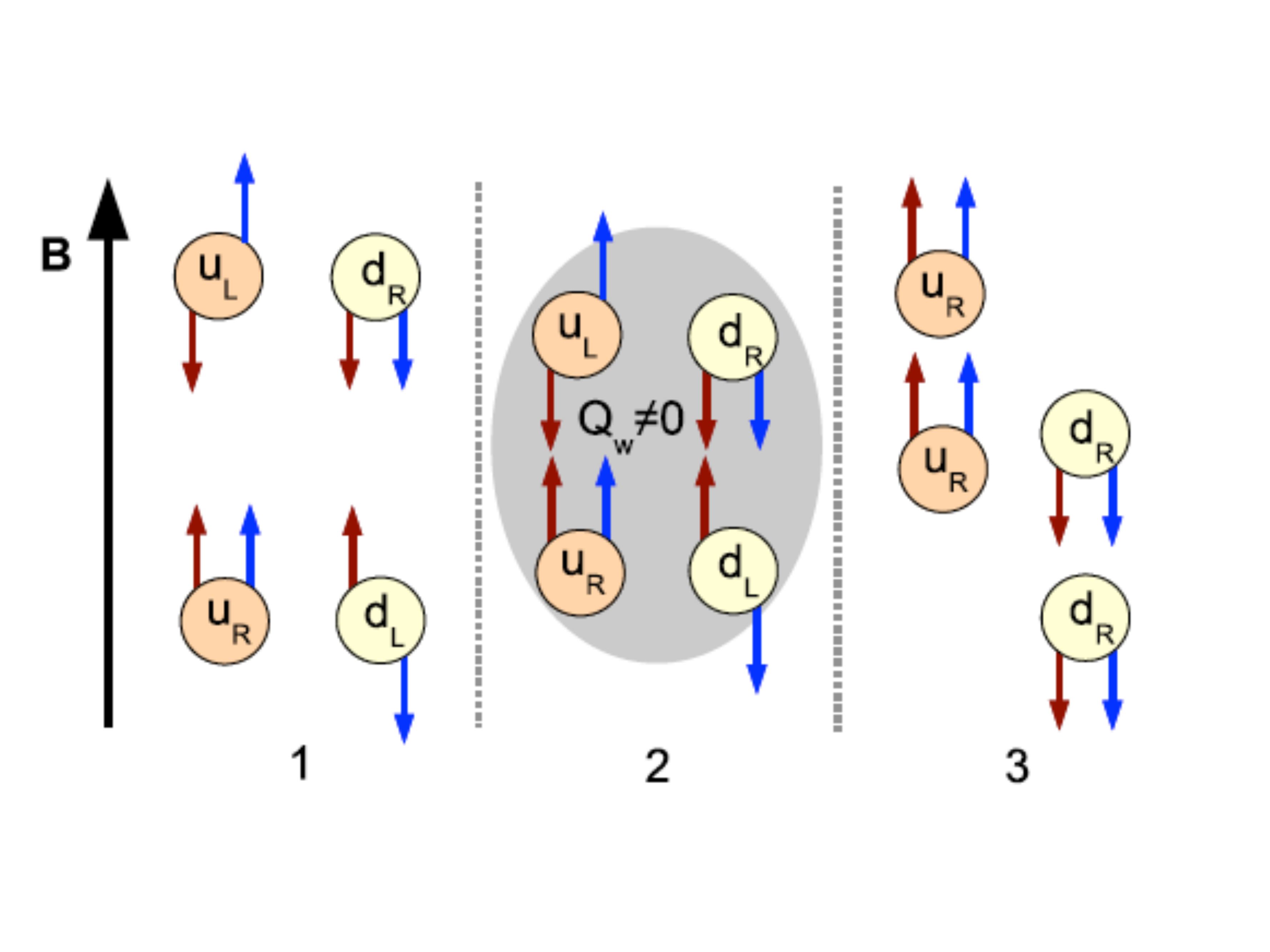}
	\caption{
A qualitative picture of the CME, from \cite{Kharzeev:2007jp}. The red arrows denote the direction of momentum, the blue arrows Ð the
spin of the quarks. (1) In a strong magnetic field, the up and down quarks
are all in the lowest Landau level and can only move along the direction of the magnetic field. Initially there are as many left-handed as right-handed quarks. (2) The
quarks interact with a gauge conÞguration with nonzero winding number $Q_w$. Assuming $Q_w = -1$,
this will convert a left-handed up/down quark into a right-handed up/down quark
by reversing the direction of momentum. (3) The right-handed up quarks will move
upwards, the right-handed down quarks will move downwards. A charge difference
of $Q = 2e$ will be created between two sides of a plane perpendicular to the magnetic
field.
 }
\label{sphaleron}
\end{figure}

In 2007, I started discussing physics with a new Brookhaven postdoc -- Harmen Warringa -- who just arrived from Amsterdam 
strongly recommended by his thesis advisor Daniel Boer, a good friend of mine. I told Harmen about my passion of the past few years -- the search for parity-odd effects in QCD matter -- and he expressed interest in collaborating on this topic. We set the goal of developing a semi-quantitative picture of the charge separation that would allow us to make predictions for the experiment. 
Larry McLerran had joined us shortly afterwards, and we began to work. 
We assumed that the chirality imbalance was due to the sphalerons in the quark-gluon plasma, and argued that the magnetic flux through the sphaleron led to the separation of electric charge. 
This phenomenon allowed a very simple intuitive explanation  \cite{Kharzeev:2007jp}, as illustrated in Fig.\ref{sphaleron} -- strong magnetic field pins down the spins of positive and negative fermions in opposite directions. This is because the lowest Landau level for chiral fermions that has zero energy is not degenerate in spin. So for magnetic field pointing upwards, the positive fermion (say, up quark) moving up and the negative fermion (say, down quark) are both right-handed with $\vec \sigma \cdot \vec p > 0$. Usually, the system possesses an equal number of left- and right-handed fermions, and hence the electric currents created by left- and right-handed fermions cancel out. 
However, if the background is topological and favors one chirality over the other, the system does develop an electric current.
\vskip0.3cm
This picture extended the mechanism of quark--antiquark asymmetry generation in \cite{Kharzeev:2004ey} by replacing the angular momentum by magnetic field. Unlike the angular momentum of the quark-gluon plasma, the magnetic field in the collision of relativistic ions can be easily computed by plugging the currents created by the initial ions and the produced charged hadrons as sources into 
the Maxwell equations. We did this, and found an extremely strong magnetic field of $eB \sim 10^{18}$ G, or a few times $m_{\pi}^2$  \cite{Kharzeev:2007jp} -- the field is so large 
at early moments in the collision because the charges of heavy ions add coherently. In such fields, the electromagnetic interactions of quarks with the field are comparable in strength to the strong interactions among the quarks! This approach allowed us to estimate the magnitude of charge asymmetry fluctuations, and to devise a number of predictions for experiment. Prompted by Larry, Harmen and I also came up with a name ``chiral magnetic effect" that we used in our paper \cite{Kharzeev:2007jp}, and later as a title of \cite{Fukushima:2008xe}.

\vskip0.3cm
However, we all knew that our paper \cite{Kharzeev:2007jp} had a number of shortcomings -- among them was the assumption of the dominance of the lowest Landau level that seemed to require a very strong (compared to the temperature of the system $T$) magnetic field $eB \gg T^2$. While the magnetic field was indeed very strong at early moments of the collision, it rapidly decreased as a function of time, and one had to find a way to i) sum over Landau levels and ii) to account for the time-dependence of magnetic field. At this moment Harmen and I were joined by Kenji Fukushima, who was a postdoc at RIKEN-BNL Center. Kenji and Harmen collaborated on the effects of magnetic field in color superconductors, so it was natural for us to join forces. After a number of discussions, we decided that a consistent way of incorporating the higher Landau levels was to evaluate the thermodynamic potential at finite chiral chemical potential, and then to differentiate it with respect to the vector potential to find the density of current $J_3 = \partial \Omega/ \partial A_3 \vert_{A_3=0}$. This computation \cite{Fukushima:2008xe} reveals the nature of the anomaly as the collective flow in Dirac sea, as discussed in section \ref{ansect} -- the current enters from the depths of Dirac sea at 
$p_3 = - \Lambda_{UV}$, and leaves our world of finite momenta at another UV boundary at  $p_3 = \Lambda_{UV}$ such that 
the UV contributions cancel out, and one is left with a final result for the CME current as explained in section {\ref{cme_sect}}. Moreover, since the excited Landau levels of chiral fermions are spin-degenerate, the only contribution to the current comes from the lowest Landau levels of zero energy -- so our assumption in \cite{Kharzeev:2007jp} in fact appeared correct for any strength of magnetic field. 
\vskip0.3cm
The fact that only the zero modes related by index theorem to the global topology contribute to the CME current highlights its topological origin. In our paper \cite{Fukushima:2008xe} we also discussed a number of other derivations, including the one using the method of Goldstone and Wilczek \cite{Goldstone:1981kk} and D'Hoker and Goldstone \cite{D'Hoker:1985yb}, and clarified the issue of the energy balance in the generation of CME current that is powered by the difference in the Fermi energies of left- and right-handed fermions. Because of this, the current cannot exist in the situations when $\mu_5$ is fixed, i.e. corresponds to a conserved chiral charge. It is the chiral anomaly that induces non-conservation of the chiral charge and thus makes the CME possible. 
\vskip0.3cm

The studies of CME on the lattice were performed in 2009 by two groups: one at ITEP that was led by Mikhail Polikarpov \cite{Buividovich:2009wi,Buividovich:2009my,Buividovich:2010tn} and another at UConn, led by Tom Blum \cite{Abramczyk:2009gb}. Both groups observed that the regions populated by topological charge in QCD matter acquire electric dipole moment in an external magnetic field. A direct measurement of the CME current on the lattice is  made possible by the observation  \cite{Fukushima:2008xe} that a finite $\mu_5$, unlike a finite baryon chemical potential, does not lead to the notorious determinant sign problem that stands in the way of lattice simulations of baryonic matter. This observation enabled the direct lattice measurement of the CME current that appeared linear in both $\mu_5$ and $B$ in accord with (\ref{chimag}) \cite{Yamamoto:2011gk,Yamamoto:2011ks}.  
\vskip0.3cm

To check that a finite $\mu_5$ does not create a problem for lattice simulations, consider the fermionic determinant in Euclidean space-time in the presence of a chiral chemical potential \cite{Fukushima:2008xe}:
\begin{equation}
 \mathrm{det} \left( \hat{D} + \mu_5 
 \gamma^0_E \gamma^5  + m \right),
\end{equation}
where $\hat{D}= \gamma_E^\mu D_\mu$, and we have chosen a
representation in which all $\gamma_E$ matrices are Hermitian,
$\gamma^0_E = \gamma^0, \gamma^i_E = i \gamma^i$.  Since
$\hat{D}$ and $\gamma_E^0 \gamma^5$ are anti-Hermitian the
eigenvalues of determinant are of the form $i \lambda_n + m$,
where $\lambda_n \in \mathbb{R}$. Because $\gamma_5$ anticommutes with
$\hat{D} + \mu_5 \gamma_E^0 \gamma^5$, all eigenvalues come in
pairs, which means that if $i \lambda_n + m$ is an eigenvalue, also
$-i \lambda_n + m$ is an eigenvalue. The determinant is the
product of all eigenvalues; thus is the product
over all $n$ of $\lambda_n^2 + m^2$.  Hence the determinant is real
and also positive semi-definite. 
\begin{figure}[t]
	\centering
	\includegraphics[width=10cm]{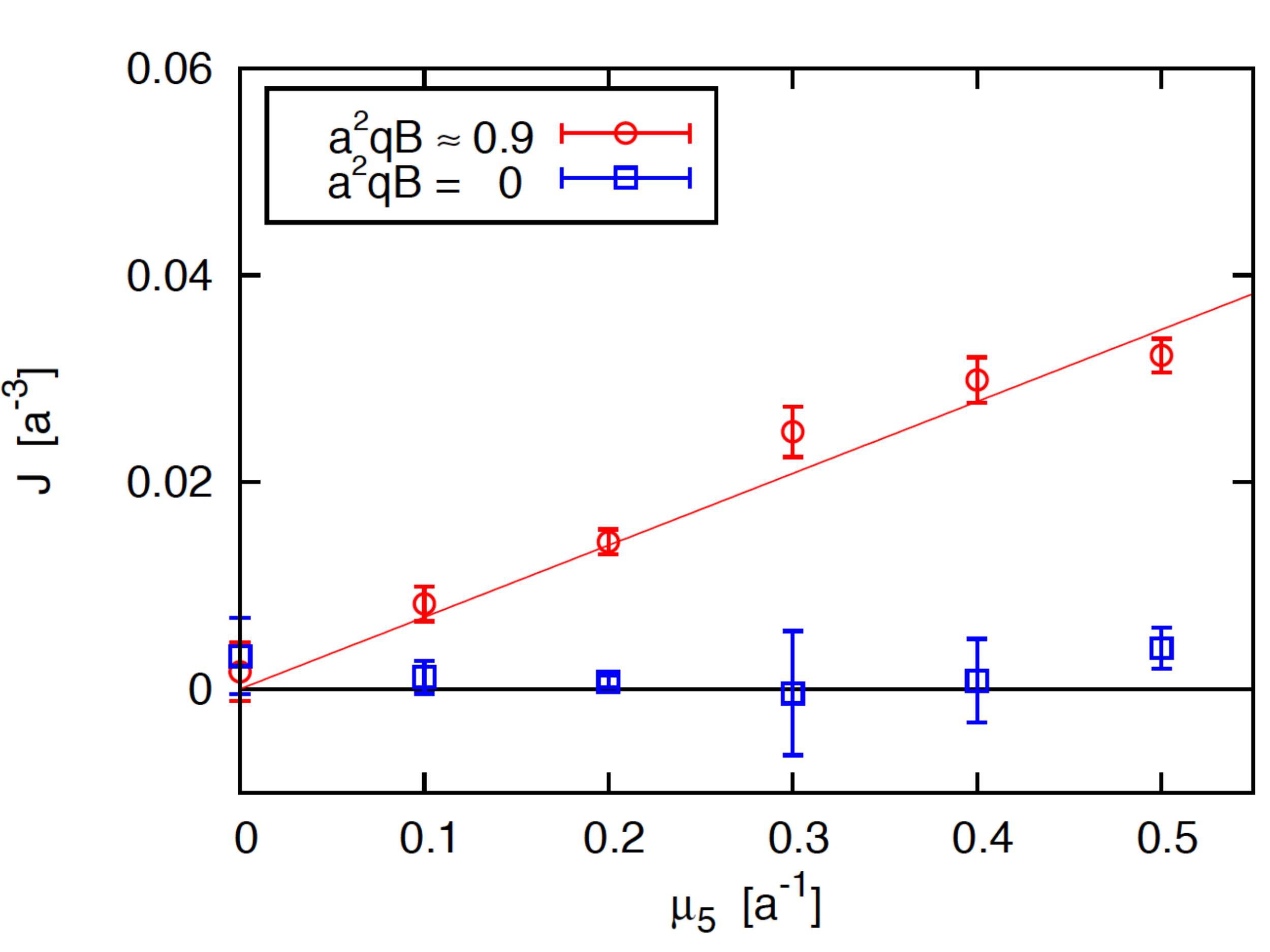}
	\caption{ The chiral magnetic current in the quark-gluon plasma as measured in full lattice QCD at $T \simeq 400$ MeV; from \cite{Yamamoto:2011ks}. The squares (blue online) indicate the absence of the current without magnetic field, and the circles (red online) -- its existence, and the expected linear dependence on the chiral chemical potential $\mu_5$, in a magnetic field. 
 }
\label{lattice}
\end{figure}

\vskip0.3cm 

The treatment of the time-dependent magnetic fields was an open issue, and Harmen and I addressed it the following year, in 2009, using the linear response theory \cite{Kharzeev:2009pj}. Unlike the zero-frequency chiral magnetic conductivity that is protected by topology, the finite-frequency response is sensitive to interactions. Indeed, one can see this by comparing the result of \cite{Kharzeev:2009pj} done in the leading order of perturbation theory to 
the result \cite{Yee:2009vw} obtained soon afterwards by Ho-Ung Yee within the holographic correspondence using the Sakai-Sugimoto model  -- at zero frequency, the results coincide, but they do differ significantly at finite frequency. We continued the studies of real-time dynamics of CME at weak coupling together with Harmen and Kenji in \cite{Fukushima:2009ft, Fukushima:2010vw}.
\vskip0.3cm

The last development from 2009 that I would like to mention in this section is the link \cite{Kharzeev:2009fn} between the CME and the Maxwell-Chern-Simons theory, or the axion electrodynamics. This allowed to make explicit the topological origin of CME deriving it through the Chern-Simons term in the effective action. The modified Maxwell-Chern-Simons equations describe in a very economical way the phenomena induced by the anomaly, including the CME and  the induction of electric charge induced on the boundary of topological domains in the presence of magnetic field \cite{Kharzeev:2009fn}. These phenomena can be described by the same theory that was used by Wilczek \cite{Wilczek:1987mv} to reproduce the Witten's effect (generation of electric charge on magnetic monopole at finite $\theta$ angle).  
\vskip0.3cm
In retrospect, I could realize the connections to axion domain walls \cite{Sikivie:1984yz} earlier had I listened more carefully to Adrian Melissinos from Rochester. Shortly after our 1998 paper  \cite{Kharzeev:1998kz} appeared, Adrian came to Brookhaven with an idea to look for the axions produced in the decay of CP-odd metastable ``bubbles" in the quark-gluon plasma. Unfortunately, the estimates showed that the  bubble was far more likely to decay into hadrons than into axions, and so the experiment never materialized -- but the link to the axion physics appeared conceptually important.
\vskip0.3cm

This concludes the ``biographic" part of the section. As time went by, I became aware of a number of important papers that preceded our work, and would like to list and discuss them here. These papers come from different fields -- astrophysics, cosmology, particle and condensed matter physics -- but address the very same problem. My only excuse for missing them earlier is that their authors were also mostly unaware of each other's work, which illustrates that the inter-disciplinary bareers still exist and have to be overcome. 

\vskip0.3cm
The first paper addressing the current generated by rotation in a chirally imbalanced system was written in 1978 by Alexander Vilenkin\footnote{I thank A. Vilenkin for kind hospitality at Tufts University and several days of enjoyable discussions on topics described in this review.}, and entitled  ``Parity Nonconservation and Rotating Black Holes'' \cite{Vilenkin:1978is}. In this, and the following papers \cite{Vilenkin:1978hb,Vilenkin:1979ui,Vilenkin:1980zv}, Vilenkin considers fermions in a rotating system and argues that the coupling of fermion's spin to the angular momentum of the system induces the current along the rotation axis. If the fermions are only left-handed (just like the Standard Model's neutrinos), then there exists a chirality imbalance that is crucial for the existence of the current. In the 1980 paper \cite{Vilenkin:1980fu} entitled ``Equilibrium Parity Violating Current In A Magnetic Field'' Vilenkin 
extends his analysis to the case of fermions in magnetic field, and derives the formula 
\beq\label{vilen}
\vec J = - \frac{e^2 \mu}{2 \pi^2} \ \vec B,
\eeq
where $\mu$ is the chemical potential of left-handed fermions. Note that this quantity has to be identified with the chiral chemical potential $\mu_5 = \mu_R - \mu_L = -\mu$ since the fermions are left-handed. The existence of a current in equilibrium was puzzling\footnote{According to A. Vilenkin, the result (\ref{vilen}) for systems in equilibrium was criticized by C.N.Yang.}, and indeed in the paper ``Cancellation Of Equilibrium Parity Violating Currents'' addressing the effect in rotating systems Vilenkin eventually concluded that ``the equilibrium current is equal to zero for all particles except neutrinos" and is exclusively the result of parity nonconservation in weak interactions. The equilibrium currents stemming from parity violation in weak interactions were then tied by Vilenkin and Leahy \cite{Vilenkin:1982pn} to the origin of cosmic magnetic fields, and were argued to produce a seed field sufficiently strong to account for the present galactic fields. 
\vskip0.3cm
According to our present understanding, the CME can exist only when the chiral chemical potential does {\it not} correspond to a conserved quantity, e.g. when
 the chiral charge is not conserved due to the chiral anomaly. Since the chiral chemical potential in this case can be viewed as the time derivative of the 
 $\theta$ angle, $\mu_5 = \partial_0 \theta$, one can see that such systems do possess an explicit dependence of a parameter describing their state  on real time and are thus strictly speaking out of equilibrium. 
 An interesting geometric prospective on this phenomenon is offered by holography, where the metric describing the systems with finite $\mu_5$ in the D3/D7 model contains a D7 brane rotating in the bulk with a finite angular velocity that is identified as $\mu_5$ in the boundary theory \cite{Hoyos:2011us,Kharzeev:2011rw}. 
\vskip0.3cm
In 1983, G. Eliashberg at
the Landau Institute pointed out the possibility of the existence of electric current induced by magnetic field in conductors with mirror isomer symmetry \cite{elia}. Even though the parity invariance in these materials is broken and thus does not forbid the relation $\vec J \sim \vec B$, eventually it was found  \cite{leonid} that such current does not exist in equilibrium\footnote{I thank L. Levitov for illuminating discussions on this topic.}, similarly to Vilenkin's conclusion in the astrophysical setup \cite{Vilenkin:1980ft}.   
\vskip0.3cm
It is instructive to re-examine the analysis made in \cite{leonid}, to see how the 
gauge invariance apparently prevents the existence of CME current in equilibrium, and to discuss how the chiral anomaly and the related topology allow it for systems with chiral imbalance\footnote{The arguments presented below were developed in a discussion with L. Levitov and H.-U. Yee.}.
Consider a conductor in magnetic field; the free energy of the system is given by
 \beq\label{freeen}
 F = \int d^3x\ \vec{j} \cdot \vec{A} - \int d^3x\ \vec{M} \cdot  \vec{B},
 \eeq
 Assume that there exists a current
 \beq
 \vec{j} = a \vec{B}.
 \eeq
Let us now use the Onsager's reciprocal relations. According to them, an off-diagonal relation between the current $\vec j$ and the field $\vec B$ implies  the off-diagonal relation  between $\vec M$ and $\vec A$:
 \beq\label{magg}
 \vec{M} = - a \vec{A}.
 \eeq
 This relation is clearly inconsistent with gauge invariance -- and since the magnetization is a physical quantity, this apparently excludes the possibility of the CME in equilibrium \cite{leonid}.  
 \vskip0.3cm
 Let us however not stop here, and continue our analysis of the free energy; from \eq{magg} and \eq{freeen} we get 
  \beq\label{free}
 F =  2 a \int d^3x\ \vec{A}\cdot \vec{B},
 \eeq
 i.e. the free energy is Chern-Simons 3-form of gauge field (``magnetic helicity") with  $a \equiv \dot{\theta} = const$. Magnetic helicity \eq{free} has a simple physical interpretation -- it counts the number of times the flux of magnetic field wraps around itself, and is thus a topological quantity. 
 The Chern-Simons form is gauge-dependent, and varies under ``large" gauge transformations that change the topology of the field $\vec A$ -- but this gauge dependence gets canceled by the appropriate boundary term. 
 \vskip0.3cm
 
 Indeed, the Chern-Simons 3-form can be viewed as emerging from the manifestly gauge-invariant ``$\theta$-term" $\theta F \tilde{F}$ with $\theta$-angle growing linearly with time, $\theta = a t$. The Chern-Pontryagin density $ F \tilde{F} $ can be represented as  a full derivative of the topological current 
 \beq
 K^\mu = \epsilon^{\mu\nu\rho\sigma} A_\nu F_{\rho\sigma},
 \eeq
 see eq(\ref{fullder}). 
 For $\theta$ linearly growing with time, $\theta = a t$ (and independent of space), we thus get
 \beq\label{thetaterm}
  \theta F \tilde{F} = \partial_\mu (\theta K^\mu) - a \ K^0,
 \eeq
 where $K^0$ yields $\vec{A} \cdot \vec{B}$ present in \eq{free}. We thus conclude that to maintain the gauge invariance, the free energy (\ref{free}) 
 has to be supplemented by the first term in \eq{thetaterm} that is a full divergence and leads to a boundary contribution. While this term, being a full divergence, does not affect the equations of motion,  it is required to restore the gauge invariance, and thus the conservation of charge. 
 \vskip0.3cm
 To see this explicitly, let us examine the effects of the boundaries that we will assume located 
 at points $z=\pm z_0$ (the magnetic field is directed along the $z$ axis). 
The integral of the boundary term is 
 \beq
 \int dt \int d^2x \ \theta \ A^0 \vec{B}\cdot \vec{n}
 \eeq   
 The corresponding electric charge density for $\theta = a t$ is obtained by differentiating w.r.t. $A^0$:
 \beq
 j_0 = a t \ \vec{B}\cdot \vec{n}
 \eeq
 This result has a simple physical meaning: at the top (bottom) surface of the sample, there is a linearly growing (decreasing) with time electric charge density -- this is a straightforward consequence of the constant in time electric current $j = a B$ propagating in the bulk. Such a surface phenomenon indeed must take place if the charge is conserved, i.e. it is a direct consequence of gauge invariance.
 \vskip0.3cm
 To summarize our discussion: the CME current through the Onsager reciprocal relations forces the free energy to be of topological Chern-Simons form. 
 This is physically admissible if the system possesses chiral fermions, since the index theorem relates the topology of the gauge field to the number of fermion zero modes. The boundary terms have to be included to restore the gauge invariance, and thus the conservation of electric charge\footnote{The argument above also helps to understand in simple physical terms the difference between the results obtained for CME from the ``consistent" and ``covariant" forms of the chiral anomaly.}.

\vskip0.3cm
In 1985, A. Redlich and L. Wijewardhana computed the induced Chern-Simons term that energes at finite temperature and/or density in a theory with an even number of left-handed $SU_L(2)$ fermion doublets \cite{Redlich:1984md}. 
They observed that a non-vanishing expectation value of the vector current emerges due to the $U_L(1)$ anomaly. The authors further speculated that the emergence of the induced Chern-Simons terms signaled an instability in the system. Also in 1985, an instability due to the induced Chern-Simons term has been pointed out by K. Tsokos \cite{Tsokos:1985ny}. An insightful paper \cite{CH} by C. Callan and J. Harvey introduced the concept of "anomaly inflow" as a physical interpretation of anomaly descent relations \cite{Zumino:1983rz}, and the corresponding anomalous currents. 
\vskip0.3cm
The 1998 paper by A. Alexeev, V. Cheianov and J. Fr{\"o}lich \cite{acf} addressed the role of chiral anomaly in the transport of charge in nano-wires, as well as in masslesss QED in magnetic field. The authors have pointed out the existence of the current induced by the chirality imbalance, related the universality of the conductivity to non-renormalization of the chiral anomaly, and discussed the relevance of gapless modes in the transport of charge. 
In the following 2000 paper \cite{Frohlich:2000en} J. Fr{\"o}lich and B. Pedrini derived the formula for the anomalous current driven by chirality imbalance from the Hamiltonian formalism and the current algebra, and addressed the relation between the chiral chemical potential and the axion field. They also discussed the anomaly-induced mechanism 
for generating primordial magnetic fields in the Early Universe. Later, the same authors have also discussed the relation between the Quantum Hall Effect, effective axion field, and the anomalous currents \cite{Frohlich:2002fg}.

\vskip0.3cm

Magnetic fields in the Universe served as a motivation also for the 1997 papers by M. Joyce and M. Shaposhnikov \cite{Joyce:1997uy}, and M. Giovannini and M. Shaposhnikov \cite{Giovannini:1997gp}. The authors considered the effect of the anomalous coupling of a primordial hypercharge magnetic field of the Standard Model to fermions, and their equations of motion contain a term describing the current proportional to the (hyper)magnetic field. In particular, the instability of the system with chiral imbalance towards the decay into (hyper)magnetic fields with non-zero Chern-Simons number has been pointed out \cite{Joyce:1997uy}. 

\vskip0.3cm
Another interesting and relevant for our discussion phenomenon is the generation of chirality through the Berry curvature in 3D condensed matter systems with 
chiral quasiparticles, so-called Weyl semimetals \cite{Abrikosov,Volovik,Wan:2011,wsm,cme}. This topic will be discussed in the next section. 

\vskip0.3cm
Finally, let me mention that the CME can be viewed as a particular kind of the classic phenomenon that was predicted by P. Curie in 1894 and named ``magnetoelectric effect" by Debye in 1926. This term refers to the coupling between magnetization and electric polarization that is exhibited for example by ${\rm Cr_2 O_3}$ and multiferroics. The origin of this effect in general is not linked to quantum anomalies, even though it is sometimes linked to chirality -- see \cite{ferro} for an example.

\section{CME: current developments}

It is impossible to list all of the recent papers on the CME and related phenomena -- for reviews dedicated to various aspects of chiral systems in magnetic field, see e.g. the recent volume \cite{Kharzeev:2012ph}. Below, I will mention some of the current research directions, and refer to representative papers.

\subsection{CME and holography}
\vskip0.2cm
\hskip 2cm {\it "The key to growth is the introduction of higher dimensions } 

\hskip 2.2cm {\it of consciousness into our awareness."} 

\vskip0.2cm

\hskip 9cm {\it Lao Tzu}

\vskip0.3cm

Because the origin of CME is topological, it is expected that the effect should survive even at strong coupling. The limit of strong coupling is accessible theoretically through the holographic correspondence, and the CME has been extensively studied in this approach, see e.g. \cite{Yee:2009vw,Rebhan:2009vc,Rubakov:2010qi,Gorsky:2010xu,Gynther:2010ed,Brits:2010pw,Hoyos:2011us,Kharzeev:2011rw,Kalaydzhyan:2011vx}, reviews \cite{Landsteiner:2012kd,Hoyos:2013qwa,Bergman:2012na,D'Hoker:2012ih} and references therein. 
\vskip0.3cm

A phenomenon similar to CME arises when instead of magnetic field there is an angular momentum  (vorticity) present -- this is so-called Chiral Vortical Effect (CVE) that within the holographic correspondence has been studied in \cite{Erdmenger:2008rm,Banerjee:2008th}. The CVE has been argued to be related to the gravitational anomaly  \cite{Landsteiner:2011cp,Landsteiner:2011iq,Jensen:2012kj,Jensen:2013kka,Jensen:2013rga}; in a holographic setup, the CVE is described through a mixed gauge-gravitational Chern-Simons term.
\vskip0.3cm

The possible relation to gravitational anomaly is very intriguing and has to be investigated further. In particular, the temperature gradient can be introduced through an effective gravitational potential \cite{Luttinger:1964zz,Stone:2012ud}, and recently it was pointed out that, through the anomaly, this leads to a new contribution to the CVE for spatially inhomogeneous systems \cite{Basar:2013qia}. Another effect emerging in the presence of inhomogeneous temperature distribution is the ``chiral heat effect" \cite{Kimura:2011ef} -- the flow of thermal current perpendicular to the gradient of temperature. 


\subsection{CME and chiral hydrodynamics}

The persistence of CME at strong coupling and small frequencies makes the hydrodynamical description of the effect possible, and indeed it emerges naturally within the relativistic hydrodynamics as shown by Son and Surowka \cite{Son:2009tf}.  The quantum anomalies in general have been found to modify hydrodynamics in a significant way, see 
\cite{Zakharov:2012vv} for a review, and refs \cite{Lublinsky:2009wr,Kalaydzhyan:2011vx,Kirilin:2012mw,Jensen:2012jy,Basar:2012bp,Fukushima:2012fg,Giovannini:2013oga,Huang:2013iia,Akamatsu:2013pjd,Baznat:2013zx,Banerjee:2013qha,Banerjee:2013fqa} for representative examples of the ongoing original work. The principle of ``no entropy production from anomalous terms" was used to constrain the relativistic conformal hydrodynamics at second order  in the derivative expansion, where it allows to compute analytically 13 out of 18 anomalous transport coefficients \cite{Kharzeev:2011ds}. 
Anomalous hydrodynamics has been found to possess novel gapless collective excitations -- the ``chiral magnetic waves" \cite{Kharzeev:2010gd}, see also \cite{Newman:2005hd}. They are analogous to sound, but in strong magnetic field propagate along the direction of the field with the velocity of light \cite{Kharzeev:2010gd}. An interesting open problem is the interplay of the CMW with the anomalous zero sound studied in \cite{Gorsky:2012gi}.
\vskip0.3cm

Anomalies have a profound importance for transport, as they make it possible to transport charges without dissipation -- this follows from the $P$-odd and $T$-even nature of the corresponding transport coefficients. The existence of CME and CVE in hydrodynamics is interesting also for the following reason -- usually, in the framework of quantum field theory one thinks about quantum anomalies as of UV phenomena arising from the regularization of loop diagrams. However, we now see that the anomalies also modify the large distance, low frequency, response of relativistic fluids. This is because the anomalies link the fermions to the global topology of gauge fields -- the much more familiar example of the IR phenomenon induced by the anomaly is the 
decay of neutral pion into two photons.

\subsection{CME and kinetic theory}

Kinetic theory has proved its usefulness in treating the approach to equilibrium. An interesting question is whether the chiral anomaly can be incorporated into kinetic description in a consistent way. This question has been addressed in a number of recent studies \cite{Pu:2010as,Son:2012wh,Gao:2012ix,Zahed:2012yu,Stephanov:2012ki,Son:2012zy,Chen:2012ca,Basar:2013iaa,Chen:2013dca}, and the answer is positive. For the kinetic description of the anomaly, it appears convenient to introduce the concept of magnetic monopole in momentum space with the corresponding Berry flux, as originally advocated by Volovik \cite{Volovik}.

\subsection{CME away from equilibrium}\label{outof}
\vskip0.2cm
A very interesting open problem is the real-time dynamics of the CME away from equilibrium; we have briefly touched upon this using the Onsager relations in section \ref{history}. These studies just begin, so we will mention just a few examples. The paper \cite{Gorsky:2010dr} addresses the decay of topological defects in magnetic field, and observes the emergence of CME current away from equilibrium. 
The self-consistent time evolution of magnetic field coupled by the anomaly to the chiral chemical potential in relativistic plasma in the presence of dissipation is analyzed in Ref.  \cite{Boyarsky:2011uy}. Another interesting example is the computation of the out-of-equilibrium CME in a holographic setup \cite{Lin:2013sga}, where the gravitational dual is the mass shell with a finite axial charge density that undergoes a gravitational collapse to a charged black hole.
\vskip0.3cm

The last example refers to the description of jet fragmentation within an effective dimensionally reduced theory  \cite{Kharzeev:2013wra} along the lines originally proposed by Casher, Kogut and Susskind \cite{Casher:1974vf}. It appears that the jet fragmentation process in this approach can be described as a propagation of the Chiral Magnetic Wave away from equilibrium; the corresponding oscillations of electric charge give rise to an intense electromagnetic radiation, possibly linking the origin of the ``anomalous soft photon puzzle" to the chiral anomaly  \cite{Kharzeev:2013wra}.

\subsection{CME in heavy ion collisions at RHIC and LHC}

As explained in section \ref{history}, the CME predicts the existence of the  
fluctuations of P-odd azimuthal asymmetries of charged hadrons that have the opposite sign for the same and opposite charge hadrons. Such fluctuations consistent with the CME have been experimentally observed at RHIC by the STAR Collaboration 
\cite{Abelev:2009uh,Abelev:2009ad}. A recent high statistics study by STAR confirms the existence of the effect, and finds that the separation of charge is predominantly orthogonal to the reaction plane \cite{Adamczyk:2013hsi}, as expected for  the CME  \cite{Kharzeev:2004ey}. 
\vskip0.3cm

The effect has been also observed at the LHC by the ALICE Collaboration \cite{Abelev:2012pa,Hori:2012hi}, with a magnitude similar to the RHIC result. Note that the similarity of the effect at RHIC and LHC energies may be simply explained by the scaling of magnetic flux through the plasma -- while the magnetic field at early times of the collision is proportional to the $\gamma$ factor, the longitudinal size of the region occupied by the produced matter shrinks as $\sim 1/\gamma$, so the magnetic flux, and thus the chiral magnetic current, are roughly independent of energy.  
\vskip0.3cm
A number of alternative explanations of the observed effect has been proposed, see e.g. \cite{Schlichting:2010qia,Pratt:2010zn,Bzdak:2012ia}. The work on the quantitative computation of the CME-induced charge asymmetries has also begun, see for example \cite{Asakawa:2010bu,Muller:2010jd,Toneev:2012zx,Toneev:2011aa,Ou:2011fm,Hongo:2013cqa}. A universal feature of all of the backgrounds to CME proposed so far is that the observed effect is attributed to a combination of the elliptic flow $v_2$ (see eq(\ref{flow}) for a definition) with a charge-dependent correlation. It is thus very important to establish whether the observed effect is driven by magnetic field or by the elliptic flow. 
\vskip0.3cm

A decisive test of this can be performed in uraniaum-uranium collisions, as proposed by Voloshin \cite{Voloshin:2010ut}. The idea is the following: the $U$ nucleus is significantly deformed, and therefore even in a central collision with no spectator nucleons the produced quark-gluon matter will be almond-shaped. The pressure gradients will thus be anisotropic, and will create a substantial elliptic flow $v_2$. On the other hand, since the collision is central and there are no spectators, the magnetic field will be close to zero. Therefore, if the effect observed in $AuAu$ collisions is due to a mundane background driven by $v_2$, it should persist in central $UU$ collisions, whereas if it is driven by magnetic field (like the CME) the effect should vanish. The first $UU$ experimental results indicate the absence of the effect in central collisions, consistent with CME 
\cite{Dong:2012mt,Wang:2012qs} -- but, needless to say, the experimental studies have to be continued.
\vskip0.3cm
There is a number of ongoing theoretical developments aimed at a quantitative understanding of the observed effect. Among important ingredients are the event-by-event fluctuations of magnetic field induced by the geometry of the collisions \cite{Bzdak:2011yy,Deng:2012pc}, the time evolution of magnetic field in the conducting medium \cite{Tuchin:2010vs,Tuchin:2013ie,Tuchin:2013apa,McLerran:2013hla}, and a quantitative theory of CME in non-stationary setups  \cite{Orlovsky:2010ga,Shevchenko:2013qya,Lin:2013sga}. 
\vskip0.3cm

In addition to the fluctuating dipole moment, the charge distribution of the quark-gluon plasma can possess the permanent quadrupole deformation induced by the Chiral Magnetic Wave (CMW)  \cite{Kharzeev:2010gd} in the presence of a vector (baryon or electric) charge density  \cite{Burnier:2011bf}. Based on this, we predict the charge dependence of the elliptic flow that is linear in the charge asymmetry \cite{Burnier:2011bf}. This prediction has been tested experimentally by STAR Collaboration  \cite{Ke:2012qb}, and the expected charge dependence has been observed. The quantitative studies of the effects induced by the CMW in a more realistic setup are underway, see e.g.  
\cite{Burnier:2012ae,Gahramanov:2012wz,Stephanov:2013tga,Taghavi:2013ena,Yee:2013cya}.

\subsection{CME in the Early Universe}

\hskip 7cm {\it "We came whirling } 

\hskip 7cm {\it out of nothingness} 

\hskip 7cm {\it scattering stars} 

\hskip 7cm {\it like dust" }\\

\hskip 7cm {\it Rumi (1207 -- 1273)}

\vskip0.3cm
A very intriguing application of CME that stimulated some of the early work \cite{Vilenkin:1982pn,Joyce:1997uy,Giovannini:1997gp,Frohlich:2000en} described above in section \ref{history} is the generation of primordial magnetic fields in the Early Universe. The chirality imbalance in the primordial electroweak plasma can be converted by the anomaly into a helical magnetic field configuration with non-zero Chern-Simons number. 
\vskip0.3cm

Among the recent developments are the transfer of magnetic helicity from small to large scales \cite{Boyarsky:2011uy}, study of the conversion of the electroweak plasma into a horizon-scale helical magnetic field \cite{Boyarsky:2012ex}, and the realization that leptogenesis can give rise to right-handed helical magnetic field that is coherent on astrophysical length scales \cite{Long:2013tha}. 
Similar ideas have been developed in the series of papers \cite{Semikoz:2007ti,Semikoz:2009ye,Semikoz:2012ka}.

The CVE has been found to lead to the production of the helical magnetic field in the turbulent electroweak plasma \cite{Tashiro:2012mf}, whereas the CME amplifies the growth of the field. \vskip0.3cm

A particularly intriguing option for the Universe is the inflation driven by a pseudo-scalar inflaton -- for example, a pseudo-Nambu-Goldstone boson \cite{Freese:1990rb} or an axion \cite{Kim:2004rp}. In this case the entire Universe would be parity-odd, and the coupling to gauge fields would induce CME currents on cosmological scales. In particular, this cosmological parity violation would be imprinted in the Cosmic Microwave Background, as discussed recently in \cite{Sorbo:2011rz,Barnaby:2011vw,Alexander:2011hz,Zhitnitsky:2013pna}. Because of an explicit out-of-equilibrium nature of inflation, this problem is naturally linked to the issues discussed in section \ref{outof}.

\subsection{CME in condensed matter physics: Weyl semimetals}

Recently, it has been realized that the triangle anomalies and the chiral magnetic effect can be realized also in a condensed matter system -- 
a (3+1)-dimensional Weyl semi-metal \cite{Wan:2011,wsm,cme}. The existence of ``substances intermediate between metals and dielectrics" with the point touchings of the valence and conduction bands in the Brillouin zone was anticipated long time ago \cite{Abrikosov}. 
\vskip0.3cm

In the vicinity of the point touching, the dispersion relation of the quasiparticles is approximately linear, as described by the Hamiltonian 
$
H = \pm v_F \vec{\sigma} \cdot \vec{k},
$ 
where $v_F$ is the Fermi velocity of the quasi-particle, $\vec{k}$ is the momentum in the first Brillouin zone, and $\vec{\sigma}$ are the Pauli matrices. 
This Hamiltonian describes massless particles with positive or negative (depending on the sign) chiralities, e.g. neutrinos, and the corresponding wave equation is known as the Weyl equation -- hence the name {\it Weyl semimetal} \cite{Wan:2011}. 

\vskip0.3cm

Weyl semimetals are closely related to 2D graphene \cite{Geim:2007}, and to the topological insulators \cite{TI,Volovik} -- 3D materials with a gapped bulk and a surface supporting gapless excitations.  Specific realizations of Weyl semimetals have been proposed, including doped silver chalcogenides $Ag_{2+\delta}Se$ and $Ag_{2+\delta}Te$ \cite{Abrikosov1}, pyrochlore irridates $A_2 Ir_2 O_7$ \cite{Wan:2011}, and 
a multilayer heterostructure composed of identical thin films of a magnetically doped 3D topological insulator, separated by ordinary-insulator spacer layers \cite{wsm}. 
Recently, the conditions for the CME in Weyl semimetals and a number of other transport phenomena induced by the anomaly were investigated in \cite{Zyuzin:2012tv,Grushin:2012mt,Kharzeev:2012dc,Son:2012bg,Goswami:2012db,Basar:2013iaa,Chen:2013,Goswami:2013bja}.

\section{Outlook}

\vskip0.2cm
\hskip 6cm {\it ``A good traveler has no fixed plans,} 

\hskip 6.2cm {\it and is not intent on arriving."} 

\vskip0.2cm
\hskip 10cm {\it Lao Tzu}
\vskip0.2cm
The CME is a spectacular example of a non-dissipative transport phenomenon induced by the chiral anomaly and protected by the global topology of the gauge field. 
This phenomenon elucidates the intricate and beautiful nature of quantum anomalies in field theory. The anomalies link quantum field theory to relativistic hydrodynamics, and induce a number of novel effects far beyond the CME.
Moreover, the anomalous CME current and related phenomena may have important practical applications since they enable the transport and processing of information and energy without dissipation. 
\vskip0.3cm
Because of this, the theory of anomalous transport deserves close attention, and should be developed further. A large number of open problems exists -- among them are the theory of real-time non-equilibrium anomalous processes, the quantitative description of charge asymmetries in heavy ion collisions, the complete theory of charge transport in Weyl semimetals and other chiral materials,  and understanding of parity violation on cosmological scales, just to name a few. Even though these problems appear to belong to different fields of science, they are tightly and intimately connected, and await courageous researchers willing to work across the boundaries of their disciplines.

\vskip0.3cm
I am grateful to my collaborators G. Basar, G. Dunne, A. Efremov, \mbox{K. Fukushima}, T. Kalaydzhyan, E. Levin, F. Loshaj, L. McLerran, R. Pisarski, M. Polikarpov, D. Son, \mbox{M. Tytgat}, R. Venugopalan, H. Warringa, \mbox{H.-U. Yee}, I. Zahed and A. Zhitnitsky for sharing their insights with me, and to A. Abanov, M. Chernodub, A. Gorsky, \mbox{U. Gursoy}, K. Jensen,  T.D. Lee, \mbox{L. Levitov}, R. Loganayagam, \mbox{A. Mazeliauskas}, V. Miransky, 
\mbox{Y. Oz}, K. Rajagopal, O. Ruchayskiy, \mbox{J. Sandweiss}, I. Shovkovy, \mbox{E. Shuryak}, M.Stephanov, \mbox{O. Teryaev}, M. Unsal, \mbox{A. Vilenkin}, S. Voloshin, {\mbox{ F. Wilczek}}, Y. Yin, and V.Zakharov for stimulating discussions. 
\vskip0.3cm

This work was supported in part by the U.S. Department of Energy under Contracts
DE-FG-88ER40388 and DE-AC02-98CH10886.





\bibliographystyle{model1a-num-names}
\bibliography{<your-bib-database>}



\end{document}